\documentclass[%
 reprint,
superscriptaddress,
preprint,
 amsmath,amssymb,
 aps,
prl,
]{revtex4-1}

\usepackage{multirow}

\usepackage{color}

\usepackage{graphicx}
\usepackage{dcolumn}
\usepackage{bm}

\begin{document}


\title{Failure of hydrogenation in protecting polycyclic aromatic
   hydrocarbons from fragmentation}

\author{M.~Gatchell}
\email{gatchell@fysik.su.se}
\affiliation{%
Department of Physics, Stockholm University, SE-106 91 Stockholm, Sweden
}%

\author{M.~H.~Stockett}%
\affiliation{%
Department of Physics, Stockholm University, SE-106 91 Stockholm, Sweden
}%
\affiliation{
 Department of Physics and Astronomy, Aarhus University, Ny Munkegade 120, DK 8000 Aarhus C, Denmark
}%

\author{N.~de Ruette}
\author{T.~Chen}
\author{L.~Giacomozzi}
\affiliation{%
Department of Physics, Stockholm University, SE-106 91 Stockholm, Sweden
}%
\author{R.~F.~Nascimento}
\affiliation{%
Department of Physics, Stockholm University, SE-106 91 Stockholm, Sweden
}%
\affiliation{Centro Federal de Educa\c c\~ao Tecnol\'ogica Celso Suckow da Fonseca, Petr\'opolis, 25620-003, RJ, Brazil}
\author{M.~Wolf}
\author{E.~K.~Anderson}
\affiliation{%
Department of Physics, Stockholm University, SE-106 91 Stockholm, Sweden
}%

\author{R.~Delaunay}
\author{V.~Vizcaino}
\author{P.~Rousseau}%
\author{L.~Adoui}%
 \affiliation{%
Centre de Recherche sur les Ions, les Mat\'eriaux et la Photonique (CIMAP),CEA-CNRS-ENSICAEN, Bd Henri Becquerel, F-14070 Caen Cedex 05, France
}%
 \affiliation{%
Universit\'e de Caen Basse-Normandie, Esplanade de la Paix, F-14032 Caen, France
}%
\author{B.~A.~Huber}%
 \affiliation{%
Centre de Recherche sur les Ions, les Mat\'eriaux et la Photonique (CIMAP),CEA-CNRS-ENSICAEN, Bd Henri Becquerel, F-14070 Caen Cedex 05, France
}%
\author{H.~T.~Schmidt}
\author{H.~Zettergren}
\author{H.~Cederquist}
\affiliation{%
Department of Physics, Stockholm University, SE-106 91 Stockholm, Sweden
}%


\begin{abstract}
A recent study of soft X-ray absorption in native and hydrogenated coronene cations, C$_{24}$H$_{12+m}^+$ $m=0\text{--}7$, led to the conclusion that additional hydrogen atoms protect (interstellar) Polycyclic Aromatic Hydrocarbon (PAH) molecules from fragmentation [Reitsma {\it et al.}, Phys.\ Rev.\ Lett.\ 113, 053002 (2014)].  The present experiment with collisions between fast (30--200 eV) He atoms and pyrene (C$_{16}$H$_{10+m}^+$, $m=0$, 6, and 16) and simulations without reference to the excitation method suggests the opposite. We find that the absolute carbon-backbone fragmentation cross section does not decrease but \emph{increases} with the degree of hydrogenation for pyrene molecules.

\end{abstract}

\pacs{Valid PACS appear here}
\maketitle



Large molecules are usually protected by the attachment of additional loosely bound atoms or molecules. Examples are bio-molecular ions in nano-droplets of water \cite{PhysRevLett.97.133401,doi:10.1021/jz3018978},	 molecules and clusters in He nano-droplets \cite{PhysRevLett.97.043201,PhysRevLett.108.076101}, clusters of fullerenes, Polycyclic Aromatic Hydrocarbons (PAHs), and/or biomolecules \cite{PhysRevLett.91.215504,AandA_442_239,doi:10.1021/jp046745z,PhysRevLett.105.213401,PhysRevA.84.043201,PhysRevA.90.022713,Delaunay:2015aa,CPHC:CPHC201000823}, and electrospray ionization where solvent molecules protect large fragile biomolecules from fragmentation \cite{Fenn06101989}. The reason for this is simple. Charge and excitation energy are rapidly redistributed over a larger system with additional internal degrees of freedom. This has been explicitly shown also for cases where the initial interactions were strongly localized \cite{PhysRevLett.91.215504,:/content/aip/journal/jcp/126/22/10.1063/1.2743433}.  

In a recent Letter, Reitsma {\it et al.}\ \cite{Reitsma:2014aa} presented a pioneering study of single and multiple H-loss following X-ray carbon K-shell ionization of hydrogenated and native cations of the PAH-molecule coronene, C$_{24}$H$_{12+m}^+$ ($m=0\text{--}7$). They observed that the coronene molecule itself (C$_{24}$H$_{12}^+$) on average lost fewer of its native H atoms when it was hydrogenated. The higher the degree of hydrogenation, the stronger was the protective effect \cite{Reitsma:2014aa}. The explanation for this was that the electronically excited coronene ions quickly evolved through Auger-emission and internal conversions into vibrationally excited ions in the electronic ground state \cite{Reitsma:2014aa}. These vibrationally excited molecules would then cool through evaporation of the extra H atoms \cite{Reitsma:2014aa}, in the same way as the evaporation of H from native PAH molecules vibrationally excited by UV photons \cite{Zhen2014211}, or by collisions with electrons \cite{Denifl2006353} or ions \cite{0004-637X-708-1-435,:/content/aip/journal/jcp/142/14/10.1063/1.4917021}. Based on this observation they concluded that the additional H atoms have a net protective effect on the PAH molecules and their carbon backbones \cite{Reitsma:2014aa}. Further, Reitsma {\it et al.}\ \cite{Reitsma:2014aa} and others \cite{doi:10.1051/0004-6361/201219952,0004-637X-752-1-3} have suggested that PAHs may be hydrogenated in interstellar space \cite{1989ApJS...71..733A,doi:10.1146/annurev.astro.46.060407.145211}. PAHs are believed to be very important for interstellar carbon chemistry and are strong candidates for carriers of so-called Diffuse Interstellar Bands \cite{doi:10.1146/annurev.astro.46.060407.145211,1989ApJS...71..733A,0004-637X-526-1-265}. The origin of these bands has long been one of astronomy's great unsolved mysteries \cite{0004-637X-526-1-265}, and only recently was C$_{60}^+$ shown to be the source for two of these spectral lines \cite{Campbell:2015aa}.

\begin{figure}[]
\includegraphics[width=0.99\columnwidth]{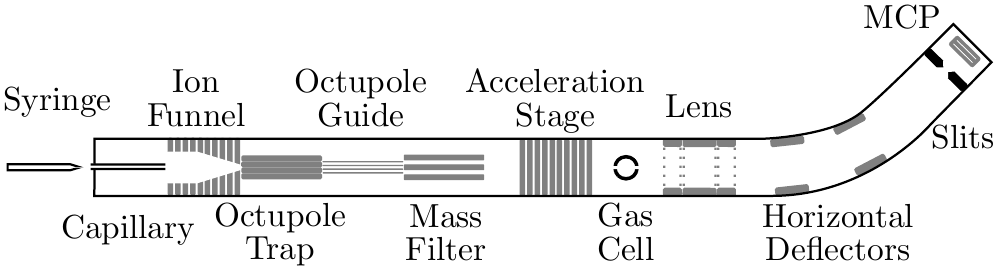}\hspace{2pc}%
\caption{Schematic of the experimental apparatus.}
\label{fig:islab}
\end{figure}

In this Letter we present experimental and simulation results which show that PAH-hydrogenation leads to \emph{larger} tendencies for carbon backbone fragmentation, at least for pyrene (C$_{16}$H$_{10}$). This is in stark contrast to the claim that hydrogenation in general has a protective effect on PAHs \cite{Reitsma:2014aa}. We have performed simulations of internally heated hydrogenated and native PAHs without reference to the excitation method that clearly show that the former loose carbon atoms more easily at a given excitation energy. We further find that pyrene cations are more easily destroyed in collisions with He atoms at center-of-mass energies of 30--200 eV when they are hydrogenated, C$_{16}$H$_{10+m}^+$ ($m=6,16$), than when they are not ($m=0$)---the larger the degree of hydrogenation, the larger the measured carbon backbone fragmentation cross section. These collision energies are typical for those found in, for instance, supernova shockwaves, where PAH molecules are processed by protons and He-ions with kinetic energies in the 10-1000 eV range \cite{Micelotta:2010aa,0004-637X-754-2-132}. The good agreement between simulations and experiment shows that the weakening of the C-C bonds due to hydrogenation is more important than the increased heat capacity and the weakening of C-H bonds following hydrogenation.

The three forms of pyrene, the native molecule and the two hydrogenated species, are purchased separately in powder form (from Sigma Aldrich), and the experiments are performed separately for each type. We produce cations of the PAH molecules from solution in an electrospray ionization source. The ions are mass-selected using a quadrupole mass filter before being accelerated by a potential that can be varied freely between 0.8 and 10 kV. The PAH ions then interact with a He target in a gas cell. The pressure in the cell is measured by means of a capacitance manometer. For these molecules this allows us to study collisions with He at center-of-mass energies of 30--200 eV. The mass-to-charge ratio of the intact ions and positively charged fragments are then determined using a series of two electrostatic horizontal deflectors and a position-sensitive microchannel plate (MCP) detector. A schematic of the experimental setup is shown in Fig.\ \ref{fig:islab} and it is described in more detail in \cite{haag2011probing}.

\begin{figure}[]
\includegraphics[width=0.99\columnwidth]{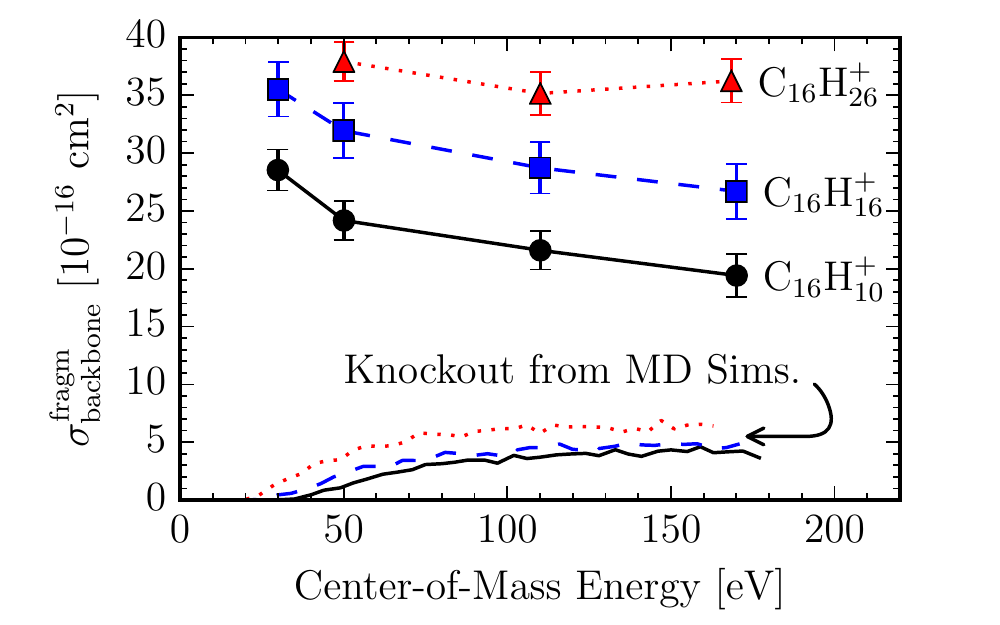}\hspace{2pc}%
\caption{Absolute cross sections for carbon backbone fragmentation, $\sigma_{\text{backbone}}^{\text{fragm}}$, in C$_{16}$H$_{10+m}^+$ + He collisions as functions of the center-of-mass collision energy. From top to bottom: Experimental results for $m=16$, $m=6$, and $m=0$ and results for prompt single- and multiple carbon knockouts from Molecular Dynamics simulations for $m=16$ (dotted curve), $m=6$ (dashed curve), and $m=0$ (solid curve) (see text).}
\label{fig:totalcs}
\end{figure}

In Fig.\ \ref{fig:totalcs} we show absolute cross sections for breaking the carbon backbone of C$_{16}$H$_{10+m}^+$ ($m=0, 6, 16$) as functions of the center-of-mass collision energy. Here, we used the beam attenuation method and at least one C-atom was lost from the PAH cation within the microsecond time-scale of the experiment. We find that the cross section \emph{increases} with increasing hydrogenation and decreasing collision energy. At the highest collision energies, the cross section for C$_{16}$H$_{26}^{+}$ is nearly twice as large as for C$_{16}$H$_{10}^{+}$ as can be seen in Fig.\ \ref{fig:totalcs}. This shows that instead of stabilizing the molecule, additional H-atoms lead to a weakening of the carbon backbone against fragmentation. The larger heat capacity of C$_{16}$H$_{26}^{+}$ and the higher rate for H-loss (lower dissociation energies, see Table \ref{tab:bindes}) obviously do not fully compensate for this weakening effect.

In order to investigate whether the increased fragmentation cross sections of the hydrogenated molecules are the result of prompt nuclear scattering processes, we have performed classical Molecular Dynamics (MD) simulations of collisions between C$_{16}$H$_{10+m}$ molecules and He. We model the molecular bonds using the reactive Tersoff potential \cite{PhysRevB.37.6991,PhysRevB.39.5566}. Nuclear scattering between He and the C- and H-atoms are described with the Ziegler-Biersack-Littmark (ZBL) potential \cite{zbl_pot_book}. The atomic coefficients used in the Tersoff potential and a more detailed description of the simulations are given in Ref.\ \cite{C4CP03293D}. In Fig.\ \ref{fig:totalcs}, we show cross sections from the simulations for prompt knockout of one or more C atoms, a process which can be important in collisions between PAH molecules and atoms/ions \cite{PhysRevA.89.032701,0004-637X-783-1-61}. The simulated knockout cross sections for C$_{16}$H$_{10}$ and C$_{16}$H$_{16}$ are almost equal and the cross section for C$_{16}$H$_{26}$ is somewhat higher due to the weakening of the C-C bonds caused by the hydrogenation (Table \ref{tab:bindes}). For all three molecules, the carbon knockout cross section accounts for less than 20\% of the measured total carbon backbone fragmentation cross section. These simulations show that statistical fragmentation of molecules vibrationally heated in the collisions is the main cause for carbon-backbone cleavage in our experiments, not knockout processes.

\begin{figure}[]
\includegraphics[width=0.95\columnwidth]{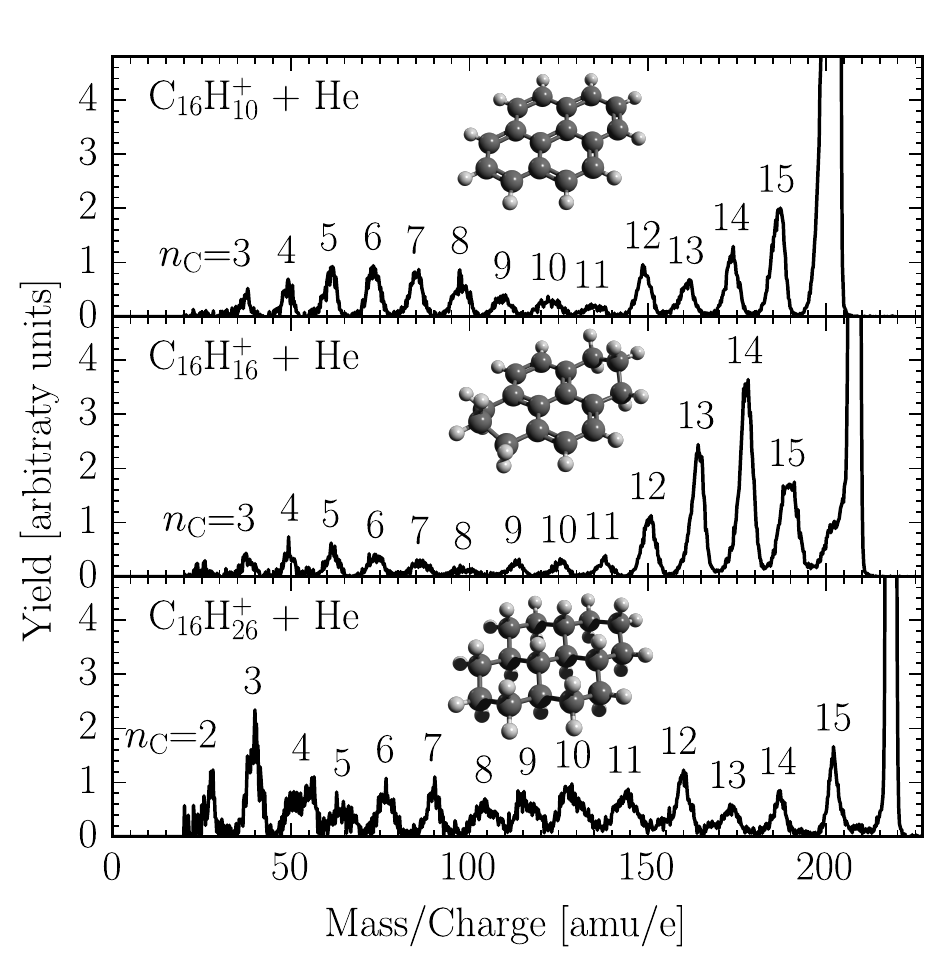}\hspace{2pc}%
\caption{Mass spectra from collisions between C$_{16}$H$_{16+m}$ ($m=0,6,16$) and He at 110 eV center-of-mass collision energy. Labels indicate the number of C atoms in every peak, $n_{\text{C}}$, each with a distribution of the number of H atoms. The insets show how the molecular structures change as H atoms are added to the pyrene molecule. The native pyrene molecule (top panel) is completely planar with only aromatic sp$^2$-type bonds. Hydrogenation results in weaker sp$^3$-type single bonds which bend the molecular structure out of the plane (middle and bottom panels).}
\label{fig:ms}
\end{figure}

Mass spectra for collisions between C$_{16}$H$_{10+m}^+$ ($m=0$, 6, 16) and He at 110 eV center-of-mass energy (by choosing the appropriate acceleration voltage for each species) are shown in Fig.\ \ref{fig:ms}. For all three molecules, the dominant feature is the peak at the mass of the intact parent molecule (202 amu, 208 amu, and 218 amu, respectively).

The upper and middle panels of Fig.\ \ref{fig:ms} show the mass spectrum for C$_{16}$H$_{10}^+$ and C$_{16}$H$_{16}^+$, respectively. While the latter has a higher total carbon backbone fragmentation cross section in collisions with He, the mass spectra for the two molecules both have somewhat similar bimodal size distributions. Important features are fragments consisting of 12--15 C atoms in both cases. In the case of C$_{16}$H$_{10}^+$, the fragments containing 15 C atoms are the result of single C knockout in direct collisions with the He target atoms \cite{PhysRevA.89.032701}. Smaller fragments are mainly due to statistical fragmentation processes, where heated C$_{16}$H$_{10}^+$ molecules, or their fragments after knockout, dissociate according to the lowest energy barriers. These secondary channels are typically associated with H- and/or C$_{2}$H$_{2}$-losses \cite{PhysRevA.85.052715}. The smallest fragments, consisting of less than about 10 C atoms, result from molecules than have been significantly heated in the collisions and fragmented through statistical channels. These small fragments are typically detected when small PAH molecules are heated, but become less important with increasing PAH size as the number of internal degrees of freedom grows \cite{PhysRevA.89.032701}. When going from C$_{16}$H$_{10}^+$ to C$_{16}$H$_{16}^+$, the yield for production of fragments containing 13 and 14 C atoms increases while that for producing fragments with less than 11 C atoms decreases (see Fig.\ \ref{fig:ms}). In both of these spectra, there are left-side tails on the main peaks (intact C$_{16}$H$_{10+m}^+$) mainly due to losses of one or a few H atoms. For C$_{16}$H$_{16}^+$, there is a more pronounced shoulder to the left of the main peak. This feature is centered at 202 amu and shows that C$_{16}$H$_{16}^+$ may lose some or all of its additional H atoms without carbon backbone fragmentation. This is a cooling effect equivalent to the one reported by Reitsma \emph{et al.}\ for hydrogenated coronene exposed to X-ray radiation \cite{Reitsma:2014aa}. However, we still measure a larger backbone fragmentation cross section for C$_{16}$H$_{16}^+$ than for C$_{16}$H$_{10}^+$ (Fig.\ \ref{fig:totalcs}) and the net effect of hydrogenation is thus still that the carbon backbone now breaks more easily.

The bottom panel of Fig.\ \ref{fig:ms} shows the mass spectrum for fully hydrogenated pyrene, C$_{16}$H$_{26}^+$. This spectrum is distinctly different from those for the molecules with fewer H atoms and has a flatter distribution biased towards smaller fragment masses. Here, the strongest peak corresponds to the C$_{3}$H$_{x}^+$ fragment. The main peak (here C$_{16}$H$_{26}^+$) is narrower than for the other molecules, indicating that pure H-loss channels are quenched by the loss of C$_n$H$_{x}$. According to our DFT calculations, the dissociation energy for CH$_{3}$-loss is indeed lower than that for H-loss for C$_{16}$H$_{26}^+$ (see Table \ref{tab:bindes}).
 
\begin{figure}[]
\includegraphics[width=0.99\columnwidth]{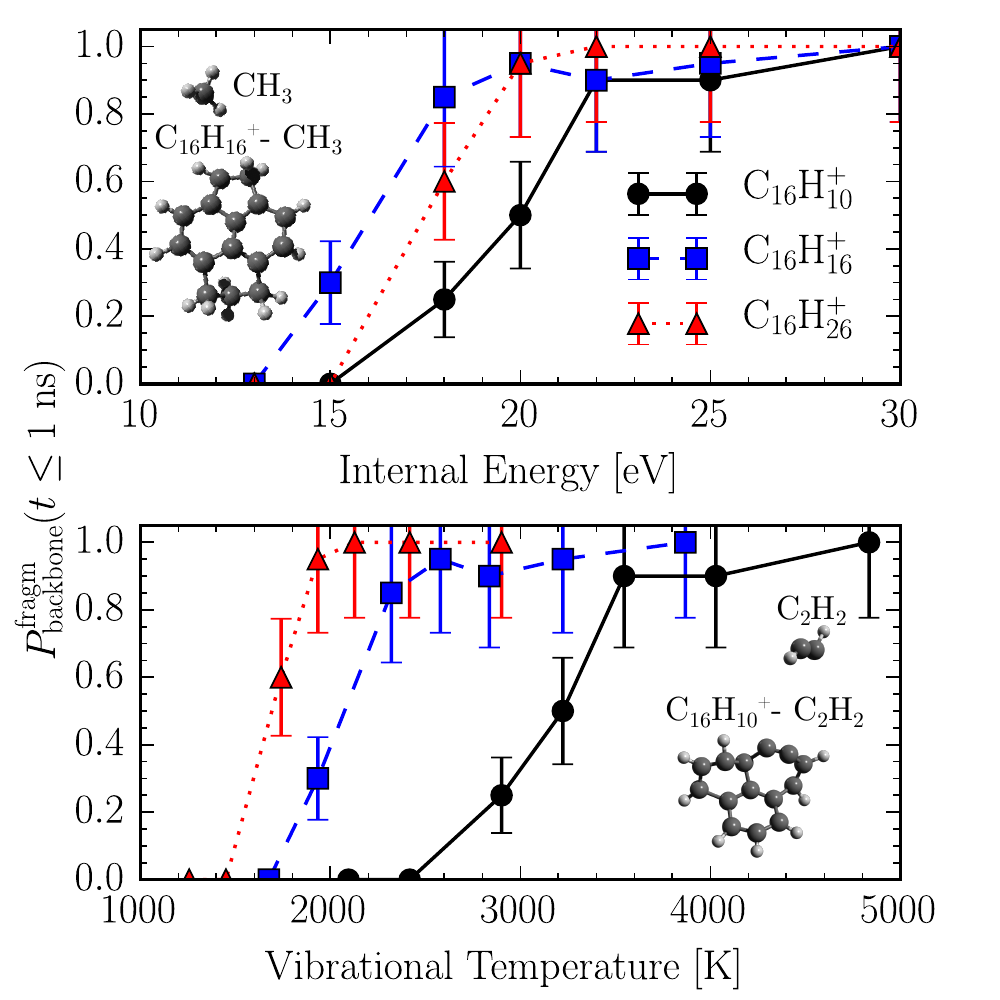}\hspace{2pc}%
\caption{Fractions of SCC-DFTB simulations which result in carbon-backbone fragmentation of C$_{16}$H$_{10+m}^+$ ($m=0,6,16$) as functions of internal energy (top panel) and vibrational temperature (bottom panel). Twenty trajectories have been simulated for each ion and each energy. The backbone fragmentation probability is higher for the hydrogenated molecules at all energies/temperatures. The insets show typical fragmentation pathways for C$_{16}$H$_{10}^+$ (C$_{2}$H$_{2}$-loss in the lower panel) and for C$_{16}$H$_{16}^+$ (CH$_{3}$-loss in the upper panel) at internal energies in the 15--20 eV range (see text).}
\label{fig:dftb}
\end{figure}

Further, we have simulated statistical fragmentation of heated C$_{16}$H$_{10+m}^+$ ($m=0$, 6, 16) ions as a function of internal energy using Self-Consistent-Charge Density-Functional Tight-Binding (SCC-DFTB) molecular dynamics \cite{PhysRevB.58.7260,Aradi:2007aa}. From these simulations, we obtain the fragmentation steps that occur within 1 nanosecond after heating the molecules to pre-set internal energies between 10 eV and 30 eV. This covers the 15--20 eV range, which is the mean energy transferred in our collisions according to the classical MD-simulations (with a tail extending up to $\sim60$ eV), and which also is the same range of vibrational energies ($\sim 20$ eV following internal conversions) as in the experiment by Reitsma \emph{et al.}\ and used in their cascade model of H evaporation from the hydrogenated PAH molecules \cite{Reitsma:2014aa,:/content/aip/journal/jcp/142/2/10.1063/1.4905471}. We have performed 20 SCC-DFTB simulations for each one of our three C$_{16}$H$_{10+m}^+$ molecules ($m=0, 6, 16$) at internal energies of 30 eV, 25 eV, 22 eV, 20 eV, 18 eV, 15 eV, and 13 eV using the mio-1-1 parameter set \cite{PhysRevB.58.7260} and DFTB+ software \cite{Aradi:2007aa}.

\begin{figure}[]
\includegraphics[width=0.99\columnwidth]{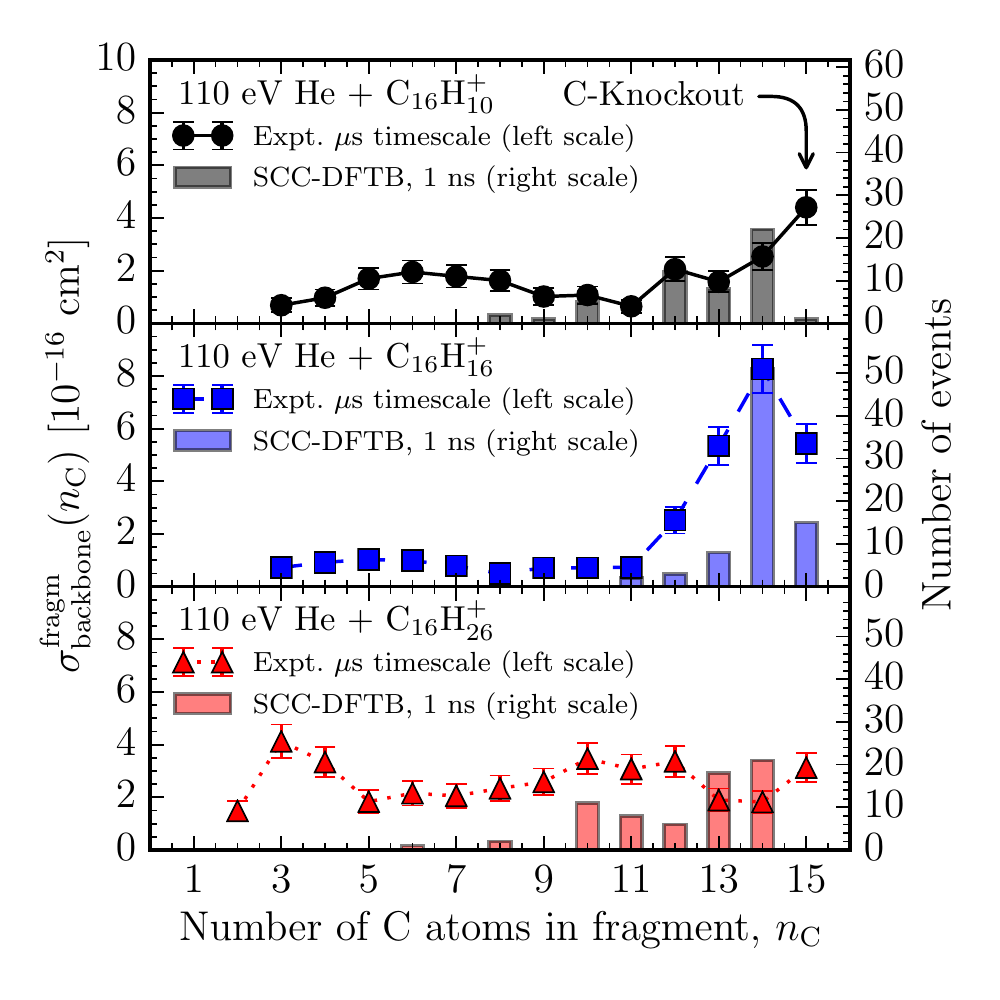}\hspace{2pc}%
\caption{Left scale: Absolute cross sections for forming products with different numbers ($n_{\text{C}}<16$) of C atoms on the microsecond timescale following 110 eV He + C$_{16}$H$_{10+m}^+$ ($m=0, 6, 16$) collisions. Right scale: The bars show the number of SCC-DFTB simulations leading to a charged fragment containing $n_{\text{C}}$ carbon atoms. We show combined result from 20 simulations (simulation time 1 ns) for each molecule for each internal energy of 15, 18, 20, 22, and 25 eV. Knockout processes are not included in these simulations, explaining the differences between experimental and SCC-DTFB results for $n_{\text{C}} = 15$.}
\label{fig:csp}
\end{figure}

The results are shown in Figs.\ \ref{fig:dftb} and \ref{fig:csp}. At internal energies above 25 eV, all three C$_{16}$H$_{10+m}^+$ molecules undergo carbon-backbone fragmentation within the 1 nanosecond SCC-DFTB simulation time. At 25 eV and below, the probability for the loss of one or more C atoms through statistical fragmentation channels is clearly higher for hydrogenated than for native pyrene (top panel of Fig.\ \ref{fig:dftb}). This difference is particularly striking when comparing the results as a function of vibrational temperature (bottom panel of Fig.\ \ref{fig:dftb}), which scales with the number of internal degrees of freedom, $3N-6$, at a given internal energy.

\begin{table}
\centering
\caption{Lowest dissociation energies (units in eV) for different fragmentation pathways calculated at the B3LYP/6-31G(d) level of theory using Gaussian09 \cite{Frisch:2009aa}.} \label{tab:bindes}
\begin{tabular}{c|c|c|c}
\toprule
Dissociation& \multirow{2}{*}{C$_{16}$H$_{10}^+$} & \multirow{2}{*}{C$_{16}$H$_{16}^+$} & \multirow{2}{*}{C$_{16}$H$_{26}^+$} \\
Channel & & & \\
\hline
H   &  5.16  & 2.56 &  2.02   \\
H + H   &    8.67   & 5.19  &  5.13   \\
CH$_{x}$       & 7.10 ($x=1$) & 2.26 ($x=3$)   &  1.60  ($x=3$)\\
C$_{2}$H$_{x}$        & 6.30 ($x=2$)  & 3.88 ($x=4$)  &  2.40 ($x=4$)  \\
C$_{3}$H$_{x}$     & 10.67 ($x=3$)  &  6.24 ($x=6$) &  2.19 ($x=6$) \\

\end{tabular}
\end{table}

The simulated fragmentation pathways (ns timescale) are shown in Fig.\ \ref{fig:csp} together with the corresponding experimental results ($\mu$s timescale). The fragmentation patterns for native pyrene ions (C$_{16}$H$_{10}^+$) follow the well established statistical pathways for PAH molecules \cite{B516437K,:/content/aip/journal/jcp/134/4/10.1063/1.3541252,PhysRevA.85.052715}. As expected we find that H-loss (6 simulations out of 100 between 15 and 25 eV) and C$_{2}$H$_{2}$-loss (20 of the 100 simulations) are significant dissociation channels. All of the remaining C$_{16}$H$_{10}^+$ dissociation pathways between 15 and 25 eV involve the loss of more than 2 C atoms, or no fragmentation. We do not observe CH$_{x}$-loss from C$_{16}$H$_{10}^+$, which is only produced by the knockout process (not included in these SCC-DFTB simulations).

For C$_{16}$H$_{16}^+$, the increased backbone fragmentation probability is due to the two carbon rings with additional H atoms. The weaker C-C bonds in these rings give rise to low dissociation energies for CH$_{3}$- and C$_{2}$H$_{4}$-loss that competes favorably with H-loss channels (Table \ref{tab:bindes}). This also results in the observed shift towards higher fragment masses for C$_{16}$H$_{16}^+$ compared to C$_{16}$H$_{10}^+$ in Fig.\ \ref{fig:csp}.

The fully saturated pyrene molecule, C$_{16}$H$_{26}^+$, has a broader fragment mass distribution than the other two molecules. The many additional H atoms in C$_{16}$H$_{26}^+$ weaken the carbon backbone such that there is no longer a single dominant fragmentation pathway. The different timescales of the simulations and experiments, the exclusion of internal energies above 30 eV, and the exclusion of knockout processes in the SCC-DFTB simulations, are the main reasons why few small fragments are seen in the simulated mass spectra.

In this Letter we have shown that hydrogenation is \emph{not} likely to protect PAH molecules from fragmentation in astrophysical environments. We have shown experimentally that hydrogenation of pyrene leads to increases in the carbon backbone fragmentation cross section in collisions with He. These results are fully supported by Self-Consistent-Charge Density-Functional Tight-Binding Molecular Dynamics simulations of vibrationally excited native and hydrogenated pyrene cations that are independent of the excitation method. This contrasts strongly to recent claims by Reitsma \emph{et al.}\ \cite{Reitsma:2014aa,:/content/aip/journal/jcp/142/2/10.1063/1.4905471} who performed measurements on hydrogenated coronene excited through carbon K-shell ionization. The question is now to what extent the size of the PAH system and/or the degree of hydrogenation matters for how well---or if at all---hydrogenation protects PAH molecules in general. For the PAH molecule pyrene, we have explicitly shown that hydrogenation \emph{does not} protect the carbon backbone of the molecule, regardless of vibrational excitation method. A simple Arrhenius picture of the competition between the increase in heat capacity and the weakening of the carbon backbone suggests that PAHs of any size will be more prone to fragmentation when hydrogenated. The reason being that backbone dissociation energies decreases faster than the heat capacity increases with additional hydrogenation. If this is the case, hydrogenation could not explain a high abundance of PAHs in the interstellar medium.

This work was supported by the Swedish Research Council (Contract No. 621-2012-3662, 621-2012-3660, and 621-2014-4501). Research was conducted in the framework of the International Associated Laboratory (LIA) Fragmentation DYNAmics of complex MOlecular systems --- DYNAMO. We acknowledge the COST actions CM1204 XUV/X-ray light and fast ions for ultrafast chemistry (XLIC) and CM0805 The Chemical Cosmos.

\bibliography{HPAH_bib}
\end{document}